\documentclass{elsarticle}
\usepackage{color}
\usepackage{array}
\usepackage{multirow}
\usepackage{multicol}
\usepackage{soul}
\usepackage{algorithm}
\usepackage{algpseudocode}

\journal{Computers, Enviroment and Urban Systems}

\algnewcommand\algorithmicinput{\textbf{Input:}}
\algnewcommand\algorithmicoutput{\textbf{Output:}}
\algnewcommand\Input{\item[\algorithmicinput]}
\algnewcommand\Output{\item[\algorithmicoutput]}

\usepackage{epsfig,pstricks,pst-plot}
\usepackage{url}
\usepackage{amssymb,amsfonts,bm,amstext, amsmath}
\usepackage{natbib}

\newcommand{\vecu}{\boldsymbol{u}}

\newcommand{\vecc}{\boldsymbol{c}}
\newcommand{\vecC}{\boldsymbol{C}}

\newcommand{\vecp}{\boldsymbol{p}}
\newcommand{\vecT}{\boldsymbol{T}}

\begin{document}
\begin{frontmatter}

\title{A Scalable Framework for Spatiotemporal Analysis of Location-based Social Media Data}

\author[ttu]{Guofeng Cao \corref{cor2}}
\author[cigi,ncsa]{Shaowen Wang\corref{cor2}}
\cortext[cor2]{Corresponding author}
\author[cigi]{Myunghwa Hwang}
\author[cigi,ncsa]{Anand Padmanabhan}
\author[cigi]{Zhenhua Zhang}
\author[cigi]{Kiumars Soltani}
\address[ttu]{Department of Geosciences, Texas Tech University, Lubbock
79409, TX, USA}
\address[cigi]{Cyberinfrastructure and Geospatial Information Laboratory, \\ 
Department of Geography and Geographic Information Science,\\
University of Illinois at Urbana-Champaign, Urbana 61801, IL, USA}
\address[ncsa]{National Center for Supercomputing Applications\\
University of Illinois at Urbana-Champaign, Urbana 61801, IL, USA}
\begin{abstract}{

In the past several years, social media (e.g., \textit{Twitter} and \textit
{Facebook}) has been experiencing a spectacular rise and popularity, and
becoming a ubiquitous discourse for content sharing and social networking.
With the widespread of mobile devices and location-based services, social
media typically allows users to share whereabouts of daily activities
(e.g., check-ins and taking photos), and thus strengthens the roles of
social media as a proxy to understand human behaviors and complex social
dynamics in geographic spaces. Unlike conventional spatiotemporal data,
this new modality of data is dynamic, massive, and typically represented
in stream of unstructured media (e.g., texts and photos), which pose
fundamental representation, modeling and computational challenges to
conventional spatiotemporal analysis and geographic information science.
In this paper, we describe a scalable computational framework to harness
massive location-based social media data for efficient and systematic
spatiotemporal data analysis. Within this framework, the concept of
space-time trajectories (or paths) is applied to represent activity
profiles of social media users. A hierarchical spatiotemporal data
model, namely a spatiotemporal data cube model, is developed based on
collections of space-time trajectories to represent the collective
dynamics of social media users across aggregation boundaries at multiple
spatiotemporal scales. The framework is implemented based upon a public
data stream of {\it Twitter} feeds posted on the continent of North
America. To demonstrate the advantages and performance of this framework,
an interactive flow mapping interface (including both single-source
and multiple-source flow mapping) is developed to allow real-time,
and interactive visual exploration of movement dynamics in massive
location-based social media at multiple scales.


} \end{abstract}


\end{frontmatter}

\section{Introduction}

Social media represents ``a group of Internet-based applications
that are built on the ideological and technological foundations of
web 2.0, and that allow the creation and exchange of user generated
content'' \citep{kaplan2010}. Typical examples include Twitter, Facebook,
Foursquare, Flickr. In recent years, these on-line applications have been
attracting hundreds of millions of users for everyday social networking
and content sharing, and at the same time collecting a huge amount of
user-generated social media data (e.g., text messages, photos, videos,
and structure of social relationship). Twitter, for example, has grown
at an exponential rate since its founding. As of December of 2013,
monthly active Twitter users have reached more than 3.9 percent of global
population and 17.9 percent of the United States, and have sent more than
300 billion of so-called $tweets$ (individual user posts)\footnote{Source:
\url{http://www.sec.gov}}. On another front, with widespread of smart
mobile devices and location-based services, location-aware mobile devices
have become prevalent access points to social media services. Accordingly,
location has become a crucial aspect of social media data. Hundreds of
millions of smartphone users carry their location-enabled smartphones
virtually every day, record and share their whereabouts and experiences
via social media. From a perspective of geographic information science
(GIScience), these users could be viewed as ubiquitous ``citizen sensors''
that move in geographic spaces, sense and share the surrounding environment
using social media contents of various kinds. The inclusion of location
or spatial dimension blurs the interface between the cyberspace of social
media and geographic space of the real world \citep{Tsou2013}, and together
with the temporal dimension, makes social media as promising proxies to
understand the social dynamics in geographic spaces.

With accesses to fine-grained social media footprints at individual
levels, location-based social media data provide a set of new lens
to examine complex social dynamics. By taking advantage of this new
modality of data source, extensive studies with significant societal
impacts have been recently reported. In behavioral sciences, for
example, massive individual geo-tagged social media records can be
used to study human activity (e.g., travel) patterns and the effects
on human life \citep{Frank2013}. By further leveraging
friendship networking information, one can quantitatively model the
patterns of human activities and then make predictions for the future
\citep[e.g.,][]{Backstrom2010,Sadilek2012a}. At an aggregate level,
a careful aggregation of social media footprints for a subpopulation
(e.g., a geographic region) could lead to a better understanding of
this subpopulation \citep{Cranshaw2012,Li2012b} and
the connections with others \citep[e.g.,][]{Wu2014}. In public health
surveillance, studies have shown that, for certain diseases (e.g.,
influenza), a careful analysis of geo-located Twitter messages could
provide surveillance capabilities comparable with the official CDC (US
Centers for Disease Control and Prevention) reports, but in a much more
timely manner \citep[e.g.,][]{Signorini2011,Nagel2013}.

The initial successes in exploiting location-based social media data,
demonstrate great potentials and provide tremendous opportunities to
gain new scientific insights. Distinct characteristics of location-based
social media data, however, pose fundamental representation, modeling
and computational challenges to GIScience, spatiotemporal databases and
spatiotemporal analysis. As described in \citep{Wang2013}, location-based
social media data generated by a massive number of social media users
are often big and produced continuously at an ever fast rate. Millions
of social media users frequently update or change their status and
locations. Consider the aforementioned case of daily new tweets and
even extend the desirable time window to months or years. Evidently,
location-based social media data and other user-generated geospatial
contents are becoming an important contribution source of big data
\citep{Manyika2011}. \citet{Gray2009} suggests a fourth paradigm, namely
data-intensive inquiries or eScience, for scientific discoveries to
survive the deluge of big data. While GIScience is shifting rapidly to
embrace the fourth paradigm \citep{Wright2011, Wang2010}, the big data
nature of location-based social media is well beyond the capability
of mainstream geographic information systems (GIS). Furthermore, the
dynamic and real-time characteristics of social media data hinder direct
applications of conventional GIS, which tends to represent the real world
as static forms instead of dynamic processes \citep{Goodchild2004d}. In
addition, social media contents are usually produced in unstructured forms
of media (e.g., texts, photos and videos) in contrast to the typical
well-structured, ready-to-use geospatial data sources. Extra efforts, such
as data retrieving and data mining processes, are often necessary to obtain
the data and then make the data meaningful and sensible.

To address these challenges, this paper presents a scalable computational
framework to harness the massive location-based social media data to
support systematic and efficient analysis of spatiotemporal dynamics. In
the presented framework, location-based social media data are firstly
regularized in terms of {\it space-time trajectories or paths} to represent
the activity profile of each social medial individual. To exploit the
unstructured contents of social media, specific data mining methods can
be plugged into the described framework to gain valuable information of
interests. As a particular example, this paper examines the chance of
influenza like illness (ILI) infection by monitoring text messages in
Twitter posts. Within the context of data warehouse and on-line analytical
processing (OLAP) \citep{Inmon2005}, a data cube model for space-time
trajectories is designed, constructed and regularly maintained to support
systematic and efficient spatiotemporal analysis of massive location-based
social media data. Specifically, this data cube frames the spatiotemporal
dynamics of location-based social media in a multidimensional space (or
a cube) of location, time and social media users, and decomposes this
multidimensional space (cube) into a multi-scale, hierarchical structure
of {\it cuboids}. A set of measures that characterize the spatiotemporal
dynamics of location-based social media are specifically defined for each
cuboid (e.g, number of social media users and activities) and each pair of
cuboids (e.g., number of travels from one cuboid to another) of the data
cube. The measures of cuboids can be flexibly merged or split according to
the dimensional intervals of interest (e.g., administrative boundaries).
Aggregation functions associated with these measures are also defined to
support data cube operations (e.g., merge and split measures of cuboids).
With the data cube model decomposed into arrays of cuboids, one can exploit
the collective spatiotemporal dynamics in particular regions of interest
at multiple levels of spatiotemporal scales (scale effects) and different
aggregation boundaries (zoning effects) in a very efficient manner. The
presented framework thus transforms the massive, dynamic and unstructured
location-based social media data into flexible geospatial datasets that
could be easily compatible with the high performance analytical environment
of cyberGIS \citep{Wang2010} and the typical work-flows of conventional GIS
analysis. Implementation details of the framework are described based on an
open access of Twitter post stream. An on-line visual analytical interface,
including single-source and multiple-source flow mapping, is developed to
allow near real-time, interactive visual exploration of multiple scales of
distribution and movement dynamics in massive location-based social media
data.

In the remainder of this paper, key concepts of data
representation, particularly space-time trajectories, are first introduced
in the Section 2. Section 3 introduces the spatiotemporal data cube model
for efficient analysis of location-based social media data. Based on a public
data stream of {\it Twitter} feeds posted on the continent of North America, the
implementation details of presented framework are discussed in Section 4.
In Section 5 the on-line flow mapping interface is introduced and
demonstrated to showcase the advantages and effectiveness of the proposed
framework. Section 6 summarizes the paper and discusses future work. 

\section{Space-time trajectories}

Consider a set of $N$ individuals frequently sharing their activities
(e.g., message posts and check-ins) through a location-based social
media platform, which exhaustively collects activities of users. To
ease the privacy and security concerns of individual users, we suppose
that the location-based social media platform is designed to collect
these data anonymously, that is, the social media platform is unaware
of the identities of individual users and no names or other personal
identifiers are shared. Each individual is assumed to move continuously in
geographic spaces, either freely in a Euclidean space, or restrictively
in a regularized network space, e.g., roads, railways, or airways, and
frequently share messages via social media channels.

The concept of \textit{space-time paths or trajectories} has long
been used as a simple and effective means for representing and
characterizing human mobility pattern \citep{Hagerstrand1970} and
spatial trajectory analysis \citep{Zheng2011}. In this paper, we assume
that each user $u_{id}(id\in[1,N])$ corresponds to a continuously
moving, lifetime space-time trajectory $T_{id}$ in a geographic
space. This ``true'' trajectory $T_{id}$ is measured and approximated
by $TS_{id}$, a series of footprints tuples of location ($s_{id}$),
timestamp ($t_{id}$) and message content ($m_{id}$) of in social media, i.e.:
$TS_{id}=\{(s_{id}^0,t_{id}^0,m_{id}^0),(s_{id}^1,t_{id}^1,m_{id}^1 ),
\ldots, (s_{id}^i,t_{id}^i,m_{id}^i), \ldots\}$, where $t_{id}^0\le
t_{id}^1 \le\ldots t_{id}^i\le\ldots$. Different from conventional
trajectories of moving objects \citep{Zheng2011} where measurements are
often abundant and sampled at regular time intervals, measurements for
trajectory of location-based social media $TS_{id}$ (i.e., user activities)
are often very temporally sparse and irregular \citep{Gao2013}. Inactive
social media users could have long time of sedative period before next
social media activities, and yet due to privacy concerns, users have
choice to disable location options when posting activities. Consequently,
the intermediate positions between measurements on $TS_{id}$ cannot be
reliably reconstructed by commonly used methods (e.g., interpolation
and map-matching) in spatial trajectory analysis. \citep{Andrienko2012}
referred to this particular type of trajectories as ``episodic movement
data''. In the following analysis, we assume that $T_{id}$ is a step
function of time $\mathcal{T}$ defined by trajectory samples $TS_{id}$,
i.e., a social media user $u_{id}$ stays at the same location during $[t_i,
t_{i+1}]$ as at $t_{i}$ until a new activity is posted at $t_{i+1}$ when
$u_{id}$ $moves$ from $s_{i}$ to $s_{i+1}$.


To characterize the massive number of space-time trajectories associated
with social media users, several geometric measures of trajectories are
particularly of interest. Individuals typically return to the same location
frequently, and the locations are ranked based on the number of visited
times in each trajectory. We refer the most frequently visited location
area as the \textit{home} of the individual. To represent the mobility
of an individual, \textit{radius of gyration} \citep{Gonzalez2008a} is
maintained based on an individual's spatial footprint. Compared with GPS
logs or mobile phone records, location-based social media data provide
access to the contents of messages or activities ($m$). Despite of being
unstructured, these contents carry important clues to latent attributes of
social media users. Specific data mining methods could be applied to derive
desirable attributes about social media users, such as health statuses
\citep{Signorini2011}, socio-demographic information \citep{Burger2011,
Rao2010}, and opinions on specific subjects \citep{OConnor2010}. As a
specific example to illustrate the proposed framework, this paper focuses
on infection spread of ILI; a previously developed text mining method
\citep{Wang2013} is applied to text messages to diagnose the chance that
a social media user is ILI affected during a time period of space-time
trajectories.

\section{A data cube model for location-based social media data analytics}

A space-time trajectory provides a representation of individual activity
footprint in the cyberspace of social media. Oftentimes, researchers
are interested in collective characteristics of a subpopulation, e.g.,
distribution of activities of a certain group of social media users at
specific regions during specific time periods. As mentioned in the previous
section, the data characteristics of location-based social media (e.g.,
massive and dynamic) hinder the application of conventional database and
analysis methods in these aggregated analysis. In the remainder of this
section, a spatiotemporal data cube model is described to support efficient
spatiotemporal analytics of aggregated statistics of massive amount of
space-time trajectories.

Data warehouse and on-line analytical processing (OLAP) were originally
designed for effective analytics of massive business transactions. In OLAP,
data are typically represented as a data cube \citep{Gray1997}, defined by
a set of \textit{fact} tables associated with a set of \textit{dimension}
tables, and \textit{hierarchies}. According to the specification of
dimension tables, a data cube discretizes a multidimensional space
into a lattice of hierarchical {\it cuboids}, with {\it base cuboids}
representing primitive compartments in the multidimensional space at the
finest level. Base cuboids are filled with values of measures specified
in the fact tables against all the dimensions. Data cube operations,
such as {\it roll-up} (merging cuboids) and {\it drill-down} (splitting
cuboids) could be applied for different levels of aggregations. The
concepts of data warehousing and data cube have been adapted to spatial
data \citep{HSK98}. Since then, an amount of efforts have been reported
to exploit the power of spatial data cube in traditional GIS and spatial
analysis \citep[e.g.,][]{Shekhar2001, Bimonte2010}, visual
analytics of spatiotemporal processes \citep{Lins2013}, analysis of
massive moving objects \citep[e.g.,][]{Orlando2007, Leonardi2014},
and closely related, analysis of mobile cyber-physical systems
\citep[e.g.,][]{Park2013,Tang2012}.

\begin{figure}
	\hspace*{-1cm}
	\scalebox{1}{\includegraphics{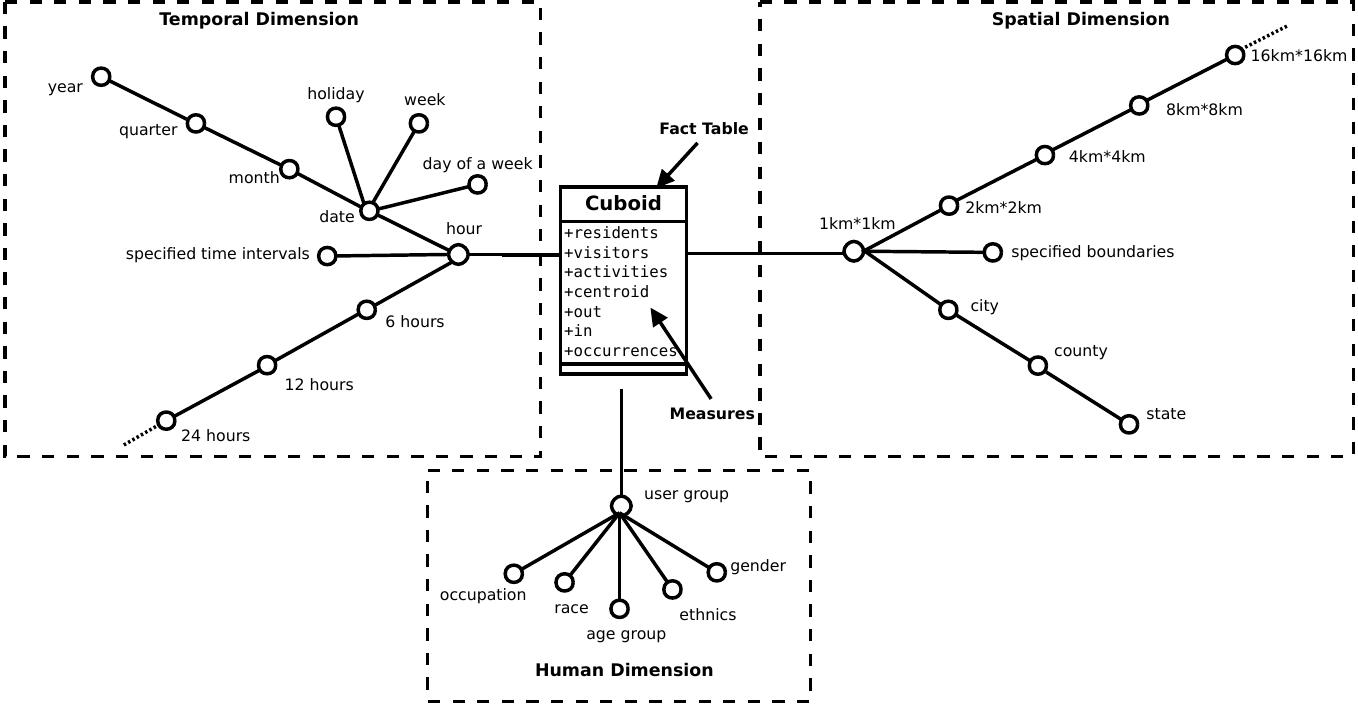}}
	\caption{Fact schema of a spatiotemporal data cube (cuboid)}
	\label{fig.cube}
\end{figure}

\begin{figure}
	\hspace*{-1cm}
	\scalebox{1}{\includegraphics{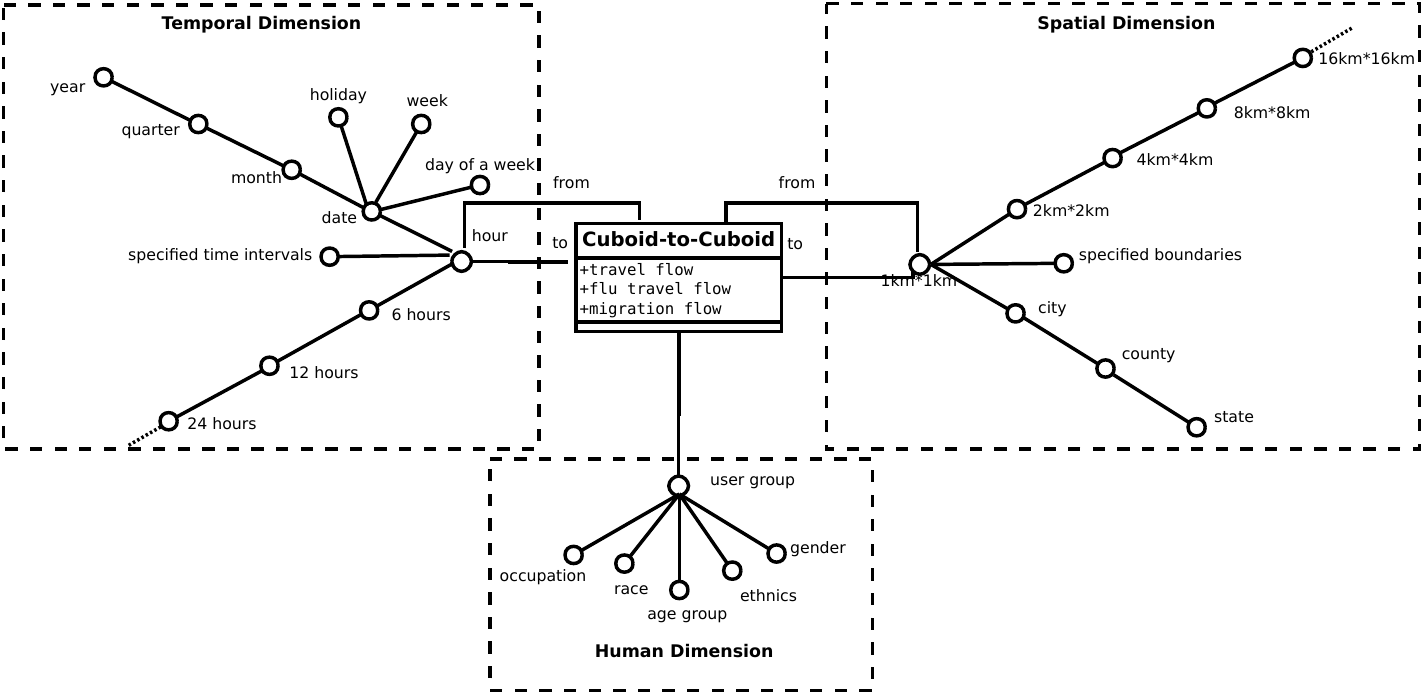}}
	\caption{Fact schema of a spatiotemporal data cube (cuboid-to-cuboid)}
	\label{fig.flow}
\end{figure}

In this paper, we extend the concept of data cube for spatiotemporal
analysis of location-based social media data. Based on the space-time
trajectories introduced previously, we are particularly interested in
investigating the human population distribution and mobility patterns
represented in location-based social media. Due to the effect of population
heterogeneity or the well known modifiable areal unit problem (MAUP)
\citep{OpenShaw1983}, conclusion drawn from a group of individuals (a
subpopulation) would probably not be applicable to another group or
aggregation of groups (population). Data cubes provide an effective tool
to examine the collective spatiotemporal dynamics of massive social media
users (space-time trajectories) at multiple levels of spatiotemporal scales
(scale effects) and different aggregation rules (zoning effects). In the
spatiotemporal data cube, we specify the dimension tables with $spatial$,
$temporal$ and $human$ (social media users) dimension. Cuboids in the data
cube are indexed by intervals at these three dimensions. To facilitate the
spatiotemporal analysis of location-based social media data, we consider
two different kinds of facts, facts for a single cuboid and facts for
interactions between pairs of cuboids. Similar as in \citet{Leonardi2014},
a graphical conceptual model for data warehouses, namely a Dimensional
Fact Model \citep{Golfarelli1998}, is adopted to represent the fact schema
within a single cuboid (Figure \ref{fig.cube}), and the fact schema between
pair of cuboids (Figure \ref{fig.flow}). For each fact schema, we introduce
a list of measures that are common and significant for characterizing
population distribution and movement dynamics represented in location-based
social media data, which will be specifically defined later in this
section. Depending on application scenarios, this data cube model is
extensible for additional appropriate measures.

Each of the three dimensions is organized at hierarchical levels of
granularities, and details of the hierarchical structures are depicted in
Figure \ref{fig.cube} and Figure \ref{fig.flow}. For the spatial dimension,
we consider a spatial grid with high resolution in the finest granularity.
In the following discussion, we choose the cell size as $1km\times 1km$
to be compatible with the commonly used global population grid datasets,
such as LandScan Global Population Database \citep{Dobson2000} and the
Global Rural-Urbana Mapping Project (GRUMP) \citep{CIESIN2004}. Different
resolution could be applied depending on application scenarios. Two
different kinds of spatial hierarchies are considered based on the
primitive spatial grid. The size of spatial grid cells keeps increasing
by merging adjacent cells in the first spatial hierarchy, and in the
other hierarchy, base grid cells aggregated by administrative boundaries
such as cities, then cities by counties, counties by states and so on.
Similarly for the temporal dimension, we choose fine temporal intervals as
base intervals. Empirically, we choose the base temporal intervals as $1$
hour in the following discussion. The base temporal intervals either keep
merging with adjacent intervals or aggregated by days, days aggregated
by weeks or months and so on. For the dimension of human (social media
users), we start from individuals, which can be further organized according
to the health status (e.g., ILI affected or not) or socio-demographic
characteristics (e.g. age groups). It should be noted that, based on base
cuboids, flexible hierarchy structures at each dimension could be defined.
In the spatial dimension, for example, base cuboids could be aggregated
according to arbitrarily specified spatial regions.


\subsection{Measures}

For a fixed group of social media users, a cuboid in the data cube
corresponds to a contiguous spatial region and a temporal interval. Given a
cuboid $\vecc$ in the data cube, the measures listed in Figure
\ref{fig.cube} can be defined for a group of users $\vecu$. For notation
simplicity, $\vecu$ is dropped in the denotation.

\begin{enumerate}
	\item ${R(\vecc)}$ (residents): the number of distinct social media users in
		$\vecu$ whose homes locate within spatial boundary of $\vecc$;
	\item ${V(\vecc)}$ (visitors): the number of distinct social media users in
		$\vecu$ who has posted activities in $\vecc$ (one user could post multiple activities);
	\item ${A(\vecc)}$ (activities): the number of social media activities by
		individuals in $\vecu$ occurring in $\vecc$;
	\item ${O(\vecc)}$ (out): the number of moves made by $\vecu$ from $\vecc$ to other cells; 
	\item ${I(\vecc)}$ (in): the number of moves made by $\vecu$ into $\vecc$ from other cells;
	\item ${S(\vecc)}$ (centroid): the expected location of social media
		activities by individuals in $\vecu$ occurring in $\vecc$;
	\item ${V_{flu}(\vecc)}$ (occurrences): the number of distinct social media users
		in $\vecu$ that post activities in $\vecc$ diagnosed as ILI
		affected occurrences; 
\end{enumerate}

Apparently, $O(\vecc)\leq V(\vecc)$, $I(\vecc)\leq V(\vecc)$, $V_{flu}(\vecc)\leq V(\vecc)$ and $V(\vecc) \leq
A(\vecc)$. Figure \ref{fig.flow} defines a list of measures quantifying the
interactions between pairs of cuboids $\vecc_i$ and $\vecc_j$:
\begin{enumerate}
	\item ${F(\vecc_i,\vecc_j)}$ (travel flows): the number of $moves$ made by social media
		users $\vecu$ starting from cuboid $\vecc_i$ and ending in
		cuboid $\vecc_j$;
	\item $F_{flu}(\vecc_i, \vecc_j)$ (flu travel flows): the number of
		$moves$
		made by social media users $\vecu$ starting from cuboid
		$\vecc_i$ and ending in cuboid $\vecc_j$ made by ILI occurrences;
	\item $F_{migration}(\vecc_i, \vecc_j)$ (migration flows): the number of 
		social media users in $\vecu$ migrating home location from
		cuboid $\vecc_i$ to cuboid $\vecc_j$;
\end{enumerate}
A $move$ will be flagged as a ILI-related one if either the starting or ending social
media activity is diagnosed as an ILI activity by the text mining method
\citep{Wang2013}. Apparently, $F_{flu}(\vecc_i, \vecc_j)\leq F(\vecc_i,\vecc_j)$ and
$F_{migration}(\vecc_i, \vecc_j)\leq F(\vecc_i,\vecc_j)$. All of these three
flow measures are asymmetry, i.e., $F(\vecc_i,\vecc_j)\neq F(\vecc_j,\vecc_i)$,
$F_{flu}(\vecc_i,\vecc_j)\neq F_{flu}(\vecc_j,\vecc_i)$,
$F_{migration}(\vecc_i,\vecc_j)\neq F_{migration}(\vecc_j,\vecc_i)$. We assume
that ${F(\vecc_i,\vecc_j)}=0$, ${F_{flu}(\vecc_i,\vecc_j)}=0$ and
${F_{migration}(\vecc_i,\vecc_j)}=0$ when $i$ equals $j$. As mentioned in
the previous section, a space-time trajectory is essentially a collection
of step functions. Therefore, flow is an aggregation of space-time
trajectories based on points of trajectories instead of the segments
in-between. 

\subsection{Aggregation functions}

Aggregation functions of measures are critical for the construction and query
operations of date cubes \citep{Gray1997}. It allows for aggregating
measures at higher levels of
the hierarchy (super-aggregates) based on those of lower levels
(sub-aggregates). Let $\mathcal{H}$ denote a spatiotemporal hierarchy
corresponding to a group of social media users $\vecu$, and without losing
generality, let $\vecc_{1}$ and $\vecc_{2}$ be two disjoint cuboids in
$\mathcal{H}$ which can be further decomposed into $K$ disjoint sub-cuboids
$\{\vecc_{1,i}, i = 1,\ldots, K\}$ with $\vecc_{1} = \cup_{i=1}^{K}\vecc_{1,i}$, and
$\{\vecc_{2,}, j = 1,\ldots, K\}$ with $\vecc_{2}=\cup_{j}^{K}\vecc_{2,j}$. Suppose
we already have the measures for sub-cuboids ($\vecc_{1,i}$ and $\vecc_{2,j}$),
aggregation functions define how to get measures for cuboids $\vecc_{1}$ and
$\vecc_{2}$. For measures $A$ of $c_{1}$, and flow measures $F$ and
$F_{flu}$ between $\vecc_1$ and $\vecc_2$, super-aggregates can be written as
recursive functions of sub-aggregates:

\begin{equation}
	A(\vecc_1) = \sum_{i=1}^{K} A(\vecc_{1,i})
	\label{eq.a}
\end{equation}

\begin{equation}
	F(\vecc_{1},\vecc_{2}) = \sum_{i,j=1}^{i,j=K} F(\vecc_{1,i},\vecc_{2,j}) 
	\label{eq.f}
\end{equation}


\begin{equation}
	F_{flu}(\vecc_{1},\vecc_{2}) = \sum_{i,j=1}^{i,j=K} F_{flu}(\vecc_{1,i},\vecc_{2,j}) 
	\label{eq.ff}
\end{equation}

\begin{equation}
	F_{migration}(\vecc_{1},\vecc_{2}) = \sum_{i,j=1}^{i,j=K} F_{migration}(\vecc_{1,i},\vecc_{2,j}) 
	\label{eq.fm}
\end{equation}

For measure $S$, $I$ and $O$, the super-aggregates $\vecc_1$ needs the
support of
other measures. Specifically for $S$: 
\begin{equation}
	S(c_1) = \frac{1}{A(c_1)}\sum_{i=1}^{K}A(c_{1,i})S(c_{1,i}) = \frac{\sum_{i=1}^{K}A(c_{1,i})S(c_{1,i})}{\sum_{i=1}^{K}A(c_{1,i})}
	\label{eq.s}
\end{equation}

For measures $I$ and $O$ on $\vecc_1$, we need to remove the space-time
trajectories that occurred within the boundaries of $\vecc_1$ according to the
definition of $I$ and $O$. Hence, we have:

\begin{equation}
	O(\vecc_1) = \sum_{i=1}^{K} O(\vecc_{1,i}) - \sum_{i=1}^{K}\sum_{j=1}^K
	F(\vecc_{1,i},\vecc_{1,j}) 
	\label{eq.o}
\end{equation}

\begin{equation}
	I(\vecc_1) = \sum_{i=1}^{K} I(\vecc_{1,i}) - \sum_{i=1}^{K}\sum_{j=1}^K
	F(\vecc_{1,i},\vecc_{1,j})
	\label{eq.i}
\end{equation}

The super-aggregates of $A$, $F$, $F_{flu}$, and $F_{migration}$ can be computed directly
from the sub-aggregates. \citet{Gray1997} categorize aggregation functions of
such measures as $distributive$ functions. For $S$, $I$ and $O$, the
computation of super-aggregates needs help from other auxiliary variables, such
as $A$, $F$, and thus $S$, thus aggregation functions of $I$ and $O$ are
$algebraic$ functions according to \citep{Gray1997}. 


Compared to Eqs. \ref{eq.a}-\ref{eq.i}, the super-aggregates of ${R}$, $V$ and
$V_{flu}$, i.e., the distinct number of residents, social media users and
social media users diagnosed as flu occurrences in a cuboid, cannot be
obtained as a recursive function of sub-aggregates. The aggregation
function of $R$, $V$ and $V_{flu}$ are $holistic$ function
\citep{Gray1997}, since the raw space-time trajectories are needed to
compute the aggregation at all levels of scales. It is computationally
impractical, particularly considering the location-based social media data
are massive and increasing continuously. It amounts to the \textit{distinct counting
problem} in spatiotemporal database, and considerable attention has been paid
to approximate the number of distinct objects in a database with auxiliary
measures. \citep{Papadias2002} simply represented the super-aggregates of such
measures as the sums of sub-cells. \citet{Tao2004} applied a probabilistic counting
approach, namely Flajolet and Martin algorithm, to aggregate the count of
distinct objects. By following \citep{Leonardi2014}, this paper approximates the
distinct number of social media users and those diagnosed as flu occurrences as:

\begin{equation}
	V(\vecc_1) = \sum_{i=1}^{K} V(\vecc_{1,i}) - \sum_{i=1}^{K}\sum_{j=1}^K F(\vecc_{1,i},\vecc_{1,j})
	\label{eq.p}
\end{equation}
\begin{equation}
	V_{flu}(\vecc_1) = \sum_{i=1}^{K} V_{flu}(\vecc_{1,i}) - \sum_{i=1}^{K}\sum_{j=1}^K F_{flu}(\vecc_{1,i},\vecc_{1,j})
	\label{eq.m}
\end{equation}

\begin{equation}
	R(\vecc_1) = \sum_{i=1}^{K} R(\vecc_{1,i}) - \sum_{i=1}^{K}\sum_{j=1}^K
	F_{migration}(\vecc_{1,i},\vecc_{1,j}) 
	\label{eq.r}
\end{equation}

As a simple illustration of this approximation in a special case, suppose
cuboids $\vecc_1$ and $\vecc_2$ share the same spatial boundaries, and
correspond to two adjacent temporal intervals $\mathcal{T}_1$ and
$\mathcal{T}_2$. There are $N$ active social media users within the spatial
boundaries posting activities both at $\mathcal{T}_1$ and $\mathcal{T}_2$.
Therefore, $V(\vecc_1)=V(\vecc_2)= F(\vecc_1,\vecc_2)=N$. According to
Equation (\ref{eq.p}), the super-aggregates for $\vecc_1\cup \vecc_2$,
$V(\vecc_1\cup \vecc_2)=N+N-N=N$, which is obviously the case. 


%

\section{Implementation} \label{chap.impl}

\begin{figure}
	\hspace*{2cm}
	\scalebox{0.5}{\includegraphics{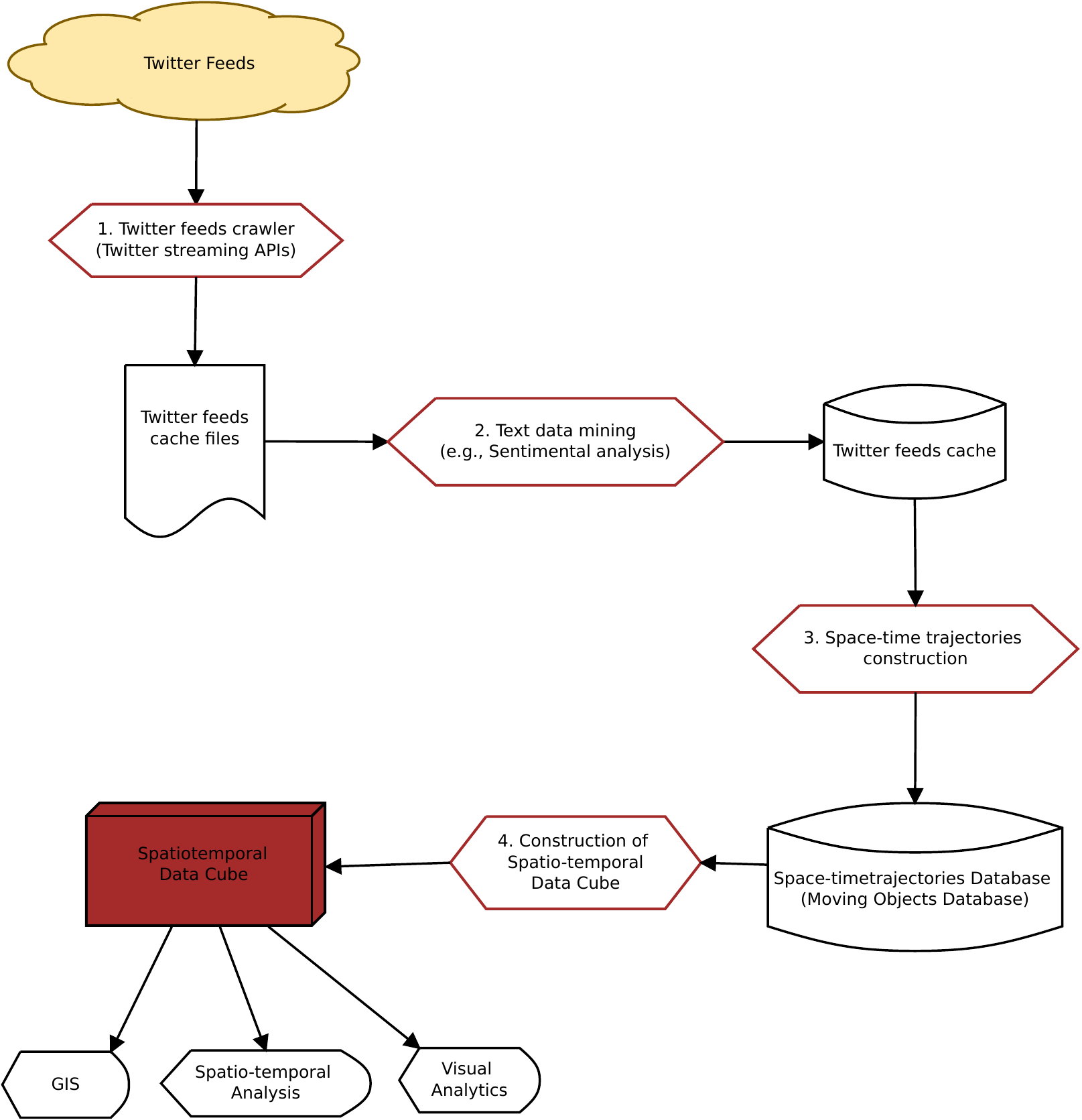}}
	\caption{Framework architecture}
	\label{fig.arch}
\end{figure}

In this section, we discuss the implementation of the data model introduced in
the previous sections based on a public data stream of Twitter feeds.
Figure \ref{fig.arch} shows the system architecture of the framework and the
data flow through different components. 

The first step is to retrieve data from Twitter. While it is millions of
social media users that are generating massive social media contents,
social media service, as hosts of these data, usually limit direct or full
access to these contents. Twitter, in particular, provides multiple levels
of interfaces to access the corpse of Twitter feeds collection. Twitter
streaming API (application programming interface), particularly, allows
anyone to near real-timely retrieve a $1\%$ sample of all the data by
specifying a set of filters, such as geographic boundaries of interests.
Despite the $1\%$ limit of sampling, it has been reported recently that
the streaming API returns almost the complete set of the geo-tagged tweets
that are of interest of this paper \citep{Morstatter2013}. A tweets crawler
was developed based on the Twitter streaming API to collect tweets posted
in the continent of North America. The returning tweets were organized as
a set of tuples $(u,s,t,m)$. In the second step, a text mining method was
applied to unstructured text messages $m$ \citep{Wang2013} to diagnose
the chance that a twitter user infected by ILI by monitoring a dictionary
of keywords related to the ILI symptoms, such as ``flu'', ``cough'',
``sneeze'' and ``fever''. It should be noted that, depending on application
scenarios, other data mining methods could be plugged in this step to gain
interested information of each tweet.

\begin{figure}
	\hspace*{1cm}
	\scalebox{0.7}{\includegraphics{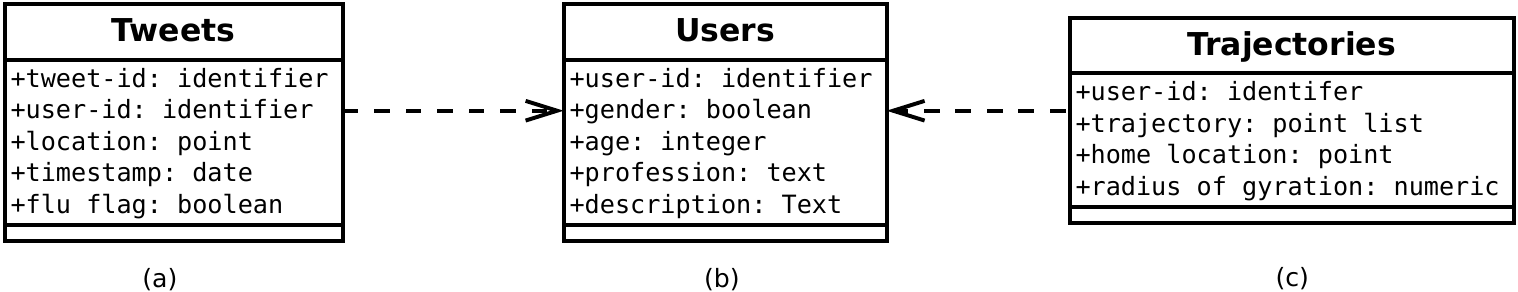}}
	\caption{Schema of a space-time trajectories database}
	\label{fig.mod}
\end{figure}

The resulted tweets are then organized into space-time trajectories
and loaded into a moving objects database \citep{Guting2005} developed
in-house based on MongoDB\footnote{http://www.mongodb.org} - a NoSQL
database. As discussed previously, we assume that each social media
user corresponds to a continuously evolving space-time trajectory to
continuously record activities. The schema of the moving objects database
is shown in Figure \ref{fig.mod}. Specifically, the {\it Users} profile
table (Figure \ref{fig.mod} (b)) describes the socio-demographic and
related information (e.g., age, gender, profession) of Tweeters indexed
by {\it user-id}. Similar with the detection of ILI infection cases, some
demographic information of Twitter users could be learned based on the
contents of tweets (e.g. \citep{Burger2011, Rao2010}). Figure \ref{fig.mod}
(a) describes the schema of resulted tweets of the preceded data mining
process, where {\it tweet-id} and {\it user-id} respectively identify a
tweet and the Twitter user ({described in the {\it Users} table}) that
posted this tweet, {\it flu-flag} indicates whether a tweet was diagnosed
as ILI infection case, {\it location} and {\it time-stamp} represent the
spatiotemporal information that the tweet was posted. Compared to the
detection of ILI infection cases, it is usually difficult to tell based on
the social media contents when the infected case would recover from the
infection. Empirically, we use an average recovery period of seven days
and flag a Twitter user, once diagnosed, as an infected case for a week.
The table of {\it Trajectories} (Figure \ref{fig.mod} (c)) describes the
life-long space-time trajectories associated with user profiles described
in the {\it Users} table. In addition to the spatiotemporal footprints
forming space-time trajectories, the geometric measures characterizing
a space-time trajectory discussed in the previous section, including
{\it home location} and {\it radius of gyration}, are also computed and
updated. Empirically, the initial {\it home location} of a space-time
trajectory corresponds to the most frequently visited place of the first
$50$ geo-located tweets activities, and keep updated afterwards. Algorithm
\ref{alg.trajectory} describes the basic procedures in the construction of
space-time trajectories.

\begin{algorithm}
	\caption{Construction of space-time trajectories}
	\label{alg.trajectory}
	\begin{algorithmic}[1]
		\Input {$\vecp$: a list of Twitter posts $p_1, p_2,\ldots$}
		\Output{$\vecT$: a set of space-time trajectories $T_{id}, id\in[1,N]$}
		\ForAll{$p_i\in \vecp$}
		\State $id \gets p_i.${\it user-id}
		\If{$T_{id}\in\vecT$}
			\State $T_{id}$.append($p_i$)
		\Else
			\State $T_{id} \gets new \mbox{ }trajectory$
			\State $T_{id}$.append($p_i$)
		  	\State $\vecT$.insert($T_{id}$)
		\EndIf
		\State update-home-location({$T_{id}$})\Comment{{\scriptsize update home location of $T_{id}$}}
		\State update-gyration-radius({$T_{id}$})\Comment{{\scriptsize update gyration radius of $T_{id}$ }}
        	\EndFor
	\end{algorithmic}
\end{algorithm}

\begin{figure}
	\hspace*{1cm}
	\scalebox{0.7}{\includegraphics{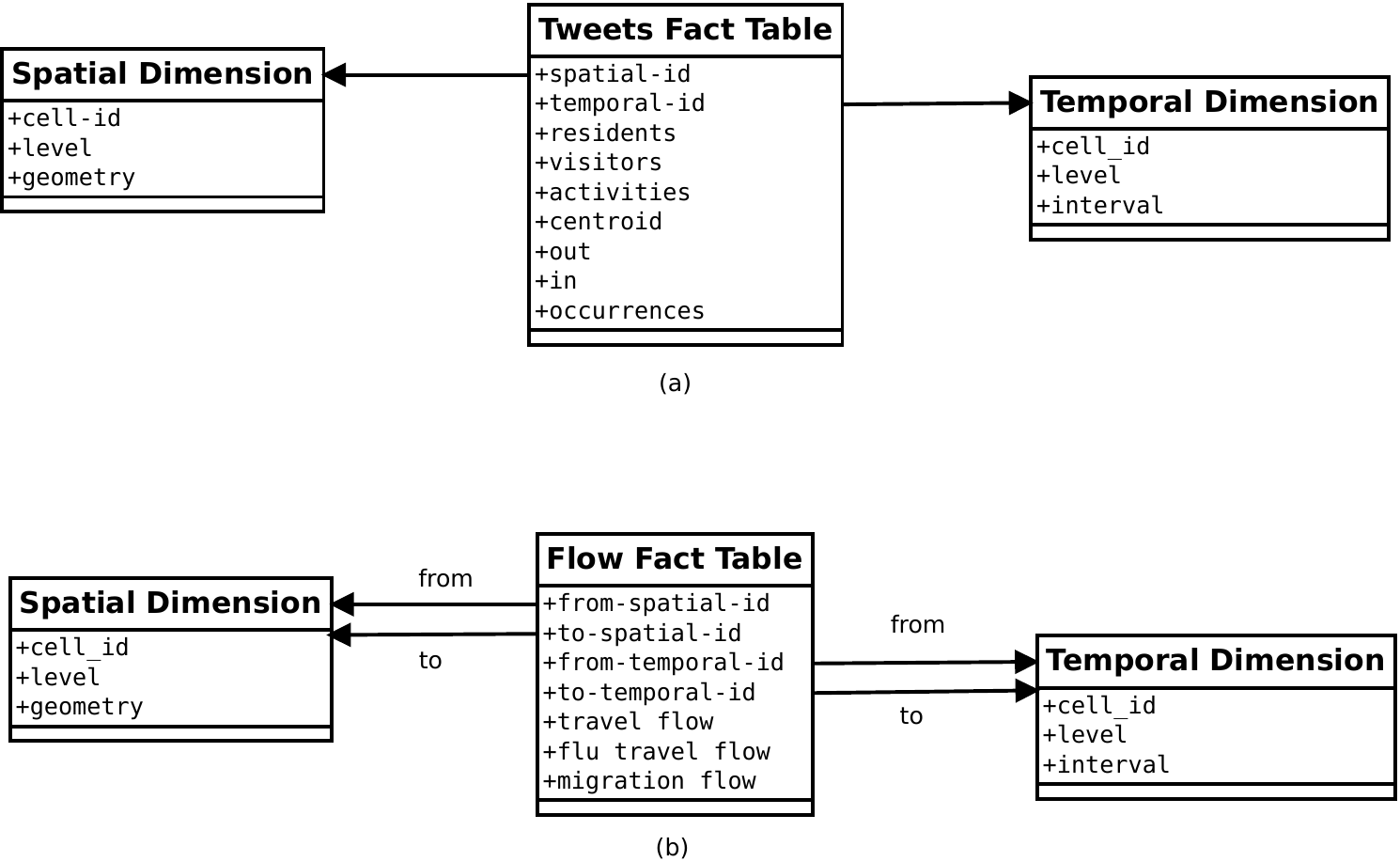}}
	\caption{Schema of a spatiotemporal data cube for Twitter feeds. There
		are two fact tables, one is for the facts of a single cuboid (a), and the
		other for facts of flows between cuboids (b).}
	\label{fig.schema}
\end{figure}

\begin{figure}
	\hspace*{-1cm}
	\scalebox{0.4}{\includegraphics{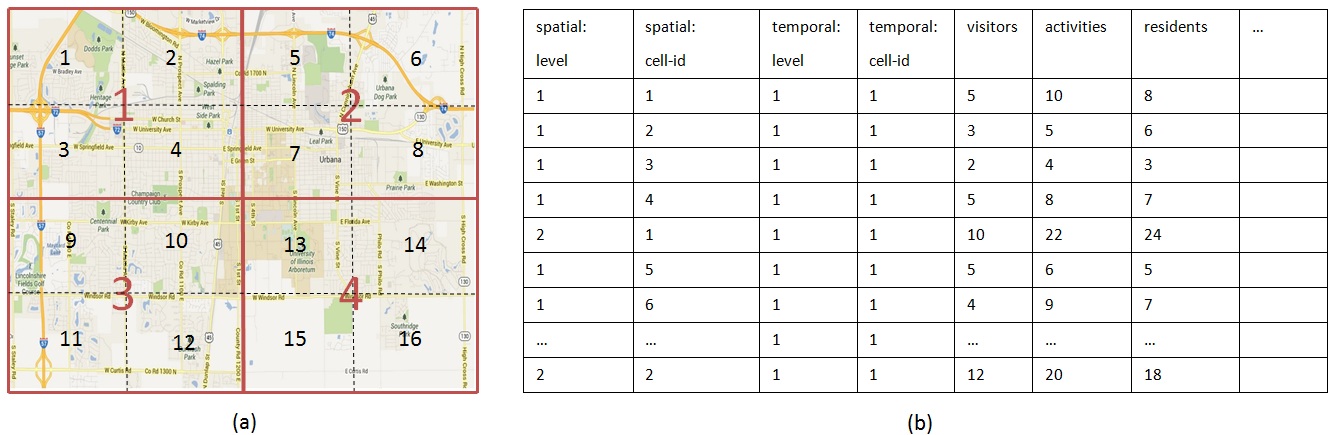}}
	\caption{An example table of a data cube with two levels of hierarchy on
	spatial dimension. (a) shows an example map and the boundaries of two
	levels of spatial hierarchy. Black dashed lines indicate the
	spatial boundaries of the cuboids at (fine) level $1$, and red solid
	lines indicate the spatial boundaries of cuboids at (coarse) level $2$; (b) is the fact table associated with each cuboid defined in (a). }
	\label{fig.example}
\end{figure}

Based on the space-time trajectories database (a moving object database)
of a group of social media users $U$, the fourth step is generating
and regularly maintaining the spatiotemporal data cube, which is
known as extract-transformation-load (ETL) process in the context
of data warehouse \citep{Inmon2005}. We consider tweets posted in the area of North America (longitude
ranging from $-167.276413^{\circ}$ to $-56.347517^{\circ}$ and latitude
from $5.499550^{\circ}$ to $82.296478^{\circ}$ in \textit{WGS84}
datum), which was divided into a lattice of cuboids with spatial cell size as
$0.008333^{\circ}$, or approximately $1km$, and temporal resolution as
{\it 1 hour} in the finest granularity (level $1$), corresponding to a
$13312\times 9216$ spatial grid evolving over time. The cell size of
the spatial grid doubles as the level (scale) increases (see Figure
\ref{fig.example}(a) for an illustration), and on the top level (level
$10$), the PPS area is decomposed into a $26\times 18$ spatial grid.
Based on the configuration of spatiotemporal data cube, Algorithm
\ref{alg.cube} can be used to compute measures for each cuboid and it
has been implemented in MongoDB. It is worth noting that function {\it
update-measures} at line 4 of Algorithm \ref{alg.cube} is a recursive
function based on Eqs. \ref{eq.a}-\ref{eq.r}. Corresponding to the
conceptual models in Figure \ref{fig.cube} and Figure \ref{fig.flow},
Figure \ref{fig.schema} shows the detail table schema for the facts of
a single cuboid (Figure \ref{fig.schema} (a)) and flows between cuboids
(Figure \ref{fig.schema} (b)). The fields of {\it level} and {\it geometry}
in the spatial dimensional table specify the structures of spatial
hierarchy, and the equivalent for the table of temporal dimension are
the fields of \textit{level} and \textit{interval}. With the temporal
interval fixed (e.g., $level = 1, interval =1$), Figure \ref{fig.example}
gives an illustrative example for the schema of a single cuboid. Figure
\ref{fig.example} (a) demonstrates two levels of spatial hierarchies, and
Figure \ref{fig.example} (b) shows the joining result of three associated
tables of spatial dimension, temporal dimension and cuboid fact tables. In
addition to the uniform spatial grids and temporal intervals, conventional
concepts of spatiotemporal databases, such as spatial join and spatial
queries, could be applied to such tables as Figure \ref{fig.example} (b),
for further aggregating the measures according to arbitrary spatiotemporal
regions.

\begin{algorithm}
	\caption{Construction of a spatiotemporal data cube}
	\label{alg.cube}
	\begin{algorithmic}[1]
		\Input{$\vecT$: a set of space-time trajectories $T_{id}, id\in[1,N]$}
		\Output{$\vecC$: a spatiotemporal data cube, or a lattice of cuboids $c_i$} 
		\ForAll{$T_i\in \vecT$}
		\State $mbb \gets$ get-mbb($T_i$) \Comment{{\scriptsize $mbb$: the minimum bounding box of $T_i$}}
		\ForAll{$c_i$ overlaps with $mbb$}
		     \State update-measures($c_i,T_i$)\Comment{{\scriptsize recursively update
		     measures according to Eqs.\ref{eq.a}-\ref{eq.r}}}
	        \EndFor	
		\EndFor
	\end{algorithmic}
\end{algorithm}

\section{An on-line mapping interface for exploration of movement dynamics in location-based social media}

The analytical framework based on the spatiotemporal data cube model frames
the massive, dynamic and unstructured location-based social media data into
a structured data model to support systematic and efficient spatiotemporal
analysis of location-based social media data. To illustrate the previously
introduced concepts and showcase the advantages of the framework, we
developed an on-line interactive visual analytical interface for the
spatiotemporal data cube. Due to the limitation of space, not all of the
concepts introduced above could be illustrated. In this section, a flow
mapping service is specifically presented based on the spatiotemporal data
cube for visual exploration of movement dynamics at multiple spatiotemporal
scales. Readers are referred to \url{http://www.flumapper.org} for further
demonstrations of the presented framework.

Flow mapping is a widely used visual analytical method to depict
and represent geographical dynamics of movement, in which each edge
represents a movement (flow) between pair-wise interacting geographical
regions. Location-based social media data provide individual-level moving
trajectories at real-time or near real-time, and thus an appealing
opportunity to investigate geographical movements, people migration in
particular, across multiple spatiotemporal scales, from macro migration
trends across the globe to characteristics of individual daily activity.
Based on the spatiotemporal data cube built for the continent of North
America, an interactive, near real-time flow mapping service for
location-based Twitter data is developed to explore the multiple scales of
flow information derived from the data cube.

\subsection{Single-source flow mapping}

Since \citet{Tobler1987} introduced a generic method to produce flow maps
with the assistance of computers, considerable efforts have been put to
improve the layout of flow maps. To remove the possible visual clutters
(e.g., edge crossings), a recent method was presented for single-source
flow mapping \citep{Verbeek2011} by taking advantage of unique features
of spiral trees. To facilitate the multi-scale visual analytics of the
flow information in the spatiotemporal data cube, this spiral tree-based
flow mapping method was adopted and implemented within an interactive
environment of cyberGIS \citep{Wang2010}. Based on the spatiotemporal
data cube model, the number of travels (i.e., flow) made by hundreds
of millions of Twitter users between pairs of specified areas during a
specified time window could be efficiently retrieved. This flow mapping
service, back-boned by the spatiotemporal data cube model, thus allows
users to interactively explore the movement dynamics of hundreds of
millions of Twitter users implied in massive location-based Twitter feeds
collection. As mentioned previously, the flu status of each Twitter users
was also learned from the contents of tweets, and it thus makes possible
that one can monitor the movement patterns of potentially flu-affected
Twitter users. In this context, this flow mapping service with support
of spatiotemporal data cube provides a promising tool for public health
researchers and practitioners.

\begin{figure}
	\hspace*{.5cm}
	\scalebox{0.3}{\includegraphics{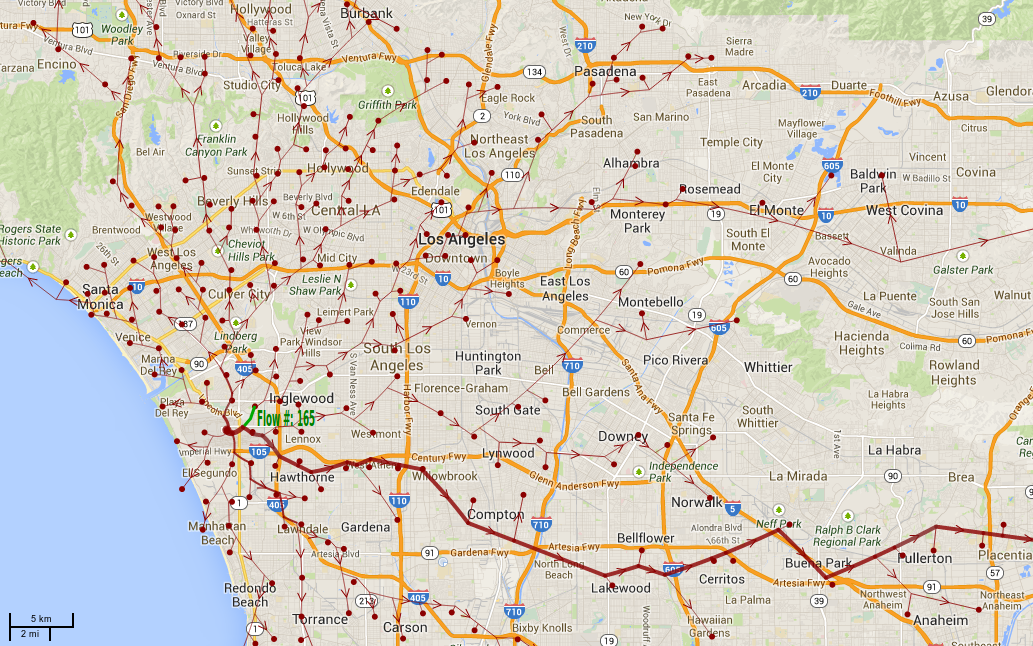}}
	\caption{A flow map of number of travels during seven days (January 29th to
	February 5th, 2014) from the LAX neighborhood to the rest area of Los Angeles.}
	\label{fig.lax2km}
\end{figure}

\begin{figure}
	\hspace*{.5cm}
	\scalebox{0.35}{\includegraphics{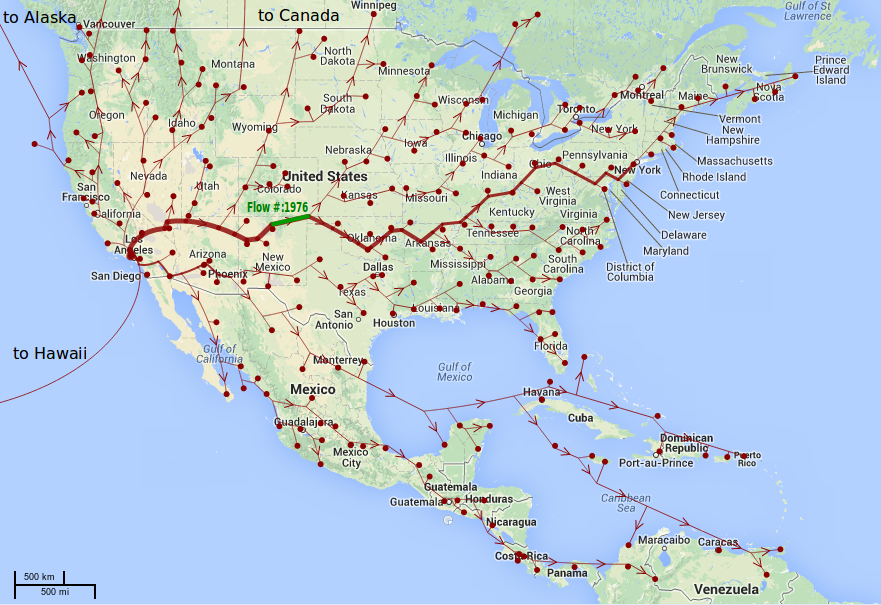}}
	\caption{A flow map of number of travels during seven days (January 29th to
	February 5th, 2014) from the Los Angeles to the other areas of
North America.}
	\label{fig.lax256km}
\end{figure}

\begin{figure}
	\hspace*{.2cm}
	\scalebox{0.4}{\includegraphics{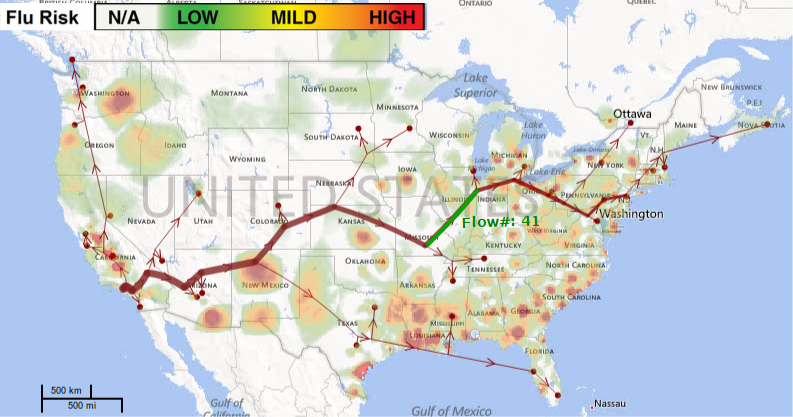}}
	\caption{A flow map of number of travels made by potential
	flu-affected Twitter users during seven days (January 29th to
February 5th, 2013) from the Los Angeles to the other areas of North
America.}
	\label{fig.lax256kmflu}
\end{figure}

As discussed in the previous section, the spatiotemporal data cube
organizes the collective dynamics of social media as hierarchical levels of
scales. Depending on the granularity of study, one could select appropriate
levels of details in the data cube. To investigate the movement dynamics in
location-based social media at a city level, for example, lower levels with
finer cell size could be more useful and for dynamics at national level,
upper levels with large cell size might be more appropriate. In the city of
Los Angeles, Figure \ref{fig.lax2km} maps seven days (16:00 January 29th,
2014 to 15:59 February 05th, 2014) of flows (number of travels in
location-based social media) that traveled out from the neighborhood of Los
Angels International airport (LAX) at the \textit{2nd} level (cell size as
$2km$) of the data cube to the other areas of the city. The solid dots
represent the centroid of each cell and the edges represents the movements
(flows) between different cells with thicker edges means more number of
flows. Users can query the exact number of flow associated with a segment
by hovering the mouse over that segment as the green segment show in Figure
(\ref{fig.lax2km} -\ref{fig.lax256kmflu}). This map with cell size of $2km$
might contain much trivial details for investigating the movement dynamics
at continental levels, in which case one could increase the aggregation
size. At the \textit{9th} granularity level of the data cube (cell size as
$256km$), Figure \ref{fig.lax256km} shows a map of flow that traveled from
the same origin and during the same period as in Figure \ref{fig.lax2km} to
the other areas of the North America.

For the same origins, Figure \ref{fig.lax256kmflu} demonstrates a flow map
of flu-affected Twitter users who traveled during the same time period as
the previous two Figures (Figure \ref{fig.lax2km} and \ref{fig.lax256km}).
One can easily check to where the potential flu-affected Twitter users
of LAX neighborhood travel. There are totally $73$ potential flu-acted
Twitter users who made travels out of the area of Los Angels during
the specified time period. As indicated in the highlighted segment in
Figure \ref{fig.lax256kmflu}, $41$ of $73$ potential flu-affected Twitter
users made travels to the northeastern area of the United States. The
background map in Figure \ref{fig.lax256kmflu} shows a risk map generated
from the occurrences of flu-affected tweets by kernel density estimation
\citep{Wang2013}. By rendering the areas with higher potential flu risk
redder and areas with lower risk greener, the background map demonstrates
near real-time distribution of flu risk across the United States. Combined
with the flow maps of travels, this mapping service could be a promising
alternative tool for the surveillance and management of flu risk at
multiple levels of spatiotemporal scales.


\subsection{Multiple-source flow mapping}

The single-source flow maps provide valuable insights into how social
media users move in and out of a particular region during a time period.
With the source location fixed, the single-source flow maps usually
lead to clean and less clutter maps. It might be difficult, however,
to glean a comprehensive view of overall move patterns of the area of
interest by considering one source at each time. The multi-source flow
mapping approach attempts to address this challenge by visualizing all
the significant movements of an area \citep{Guo2006}. As we increase the
spatial granularity level (e.g. from level $2$ to level $1$ in Figure
\ref{fig.example}), the number of cuboids and associated movement flows
increases exponentially, which tends to lead to maps with dramatic visual
clutters. To address this issue, we developed an interactive scalable
level-of-detail approach to visualize all-to-all movements of a specific
area based on the spatiotemporal data cube model.

The core part of this approach is to adaptively select the locations or
nodes (i.e., centroids of cuboids) that are critical to represent the
movement flow patterns at each level of spatial granularity. A node is
considered as critical if its score, which amounts to sum of incoming
and outgoing degree, is ranked high enough both global (among all the
nodes) and locally (among its neighbor cells). To address the scalability
issue for massive number of nodes, we implemented this filtering process
using an Apache Hadoop\footnote{http://hadoop.apache.org/} cluster to
take advantage of the distributed computation resources across multiple
computing cores. The commonly used flow mapping algorithms, namely Force
Directed Edge Bundling \citep{Holten2009}, is then applied to visualize the
flows between the selected critical nodes. This algorithm bundles the close
movements together to further reduce the clutter.

\begin{figure}
	\hspace*{.5cm}
	\scalebox{0.3}{\includegraphics{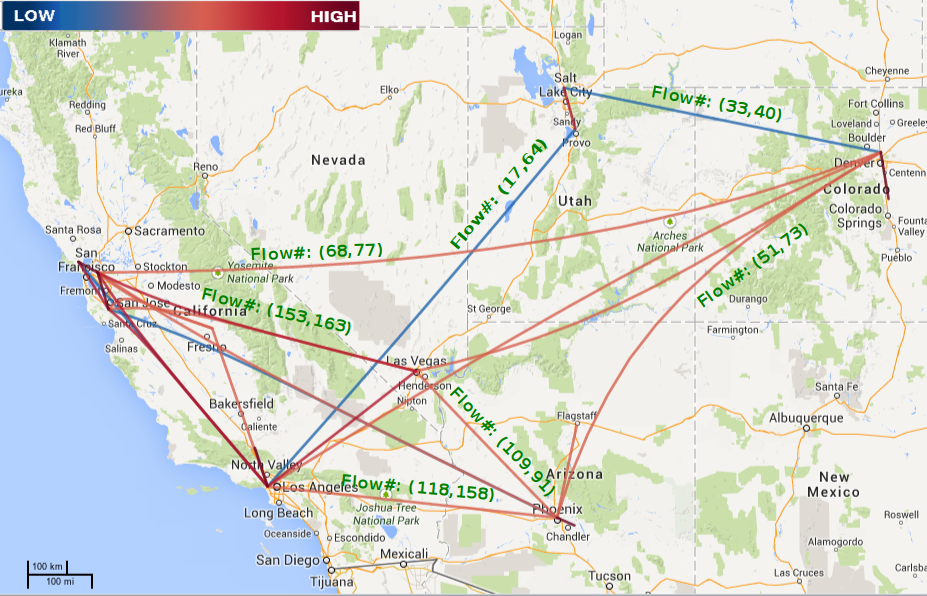}}
	\caption{Multiple-source flow maps of the travel flows between
	major cities in the southwest of the United States during the 22:00
of January 31st, 2014 to the 21:59 of February 7th, 2014.}
	\label{fig.mm-western}
\end{figure}

\begin{figure}
	\hspace*{.5cm}
	\scalebox{0.4}{\includegraphics{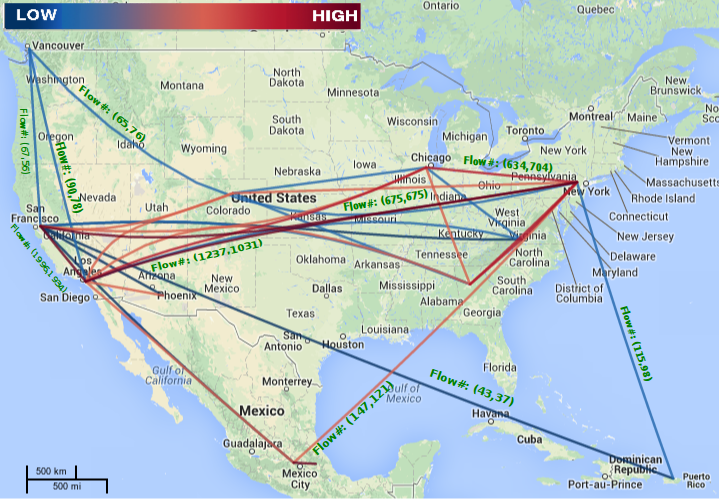}}
	\caption{Multiple-source flow maps of the travel flows of
	the major cities in North America during the 22:00
of January 31st, 2014 to the 21:59 of February 7th, 2014.}
	\label{fig.mm-na}
\end{figure}

As an example, Figure \ref{fig.mm-western} shows a result of multi-source
flow map for the movement flows at a regional scale between major cities
in the southwest of the United States (e.g., Los Angeles, San Francisco,
Las Vegas, Phoenix, Denver and Salt Lake city) during the time period
between 22:00 of January 31st, 2014 and 21:59 of February 7th, 2014.
The resulting flow maps are presented according to a coloring schema
with color of red for larger number of movement flows and color of blue
for smaller number of movement flows. It should be noted here that only
the movement flows within the extent of display window are visualized,
although the data cube is built for the entire North America. In Figure
\ref{fig.mm-western}, travels between selected pairs of cities are listed
along the connecting edges. From Los Angels to Phoenix, for example, the
listed label `Flow\#: ($118,158$)' means there were $118$ number of travels
made from Los Angels to Phoenix and $158$ for the opposite direction.
In Figure \ref{fig.mm-western}, one can clearly see the travel patterns
among the listed cities captured in the cyberspace of Twitter. During the
specified time period, it is apparent that most travels happen between the
listed cities and the associated suburb areas, and more travels between
Denver and western cities (e.g., Las Vegas, Los Angels and Phoenix) than
Denver and Salt Lake city, although Salt Lake city is significantly closer
to Denver than the western cities. Figure \ref{fig.mm-na} shows a resulting
flow map that covers the entire area of the North America during the same
time period as in Figure \ref{fig.mm-western}. Similarly, the number of
travel between cities are labeled. One can clearly see the overall travel
patterns at a continental scale between major cities in the North America
through the multiple-source flow mapping.

\section{Conclusion and future work}

In this paper, we presented a general framework to harness the massive
location-based social media data for scalable and efficient spatiotemporal
analysis of massive location-based social media data. In the presented
framework, we first adopted the concept of space-time trajectories (or
path) to represent the activities of social media users. An individual
social media user corresponds to a continuously evolving space-time
trajectory. By representing the potential path space as a lattice of
primitive cuboids, a hierarchical multi-scale model, namely a data cube
model, is designed, constructed and regularly maintained to support
systematic spatiotemporal analysis of location-based social media data.
Based on the data cube, one can easily query and summarize the
spatiotemporal distribution and dynamics in location-based social media by
specified aggregation boundaries, such as spatial regions, time duration
and population group at multiple scales. The system architectures and
implementation details based on an public Twitter feeds collection of the
Untied States were discussed. To showcase the effectiveness of the
presented framework, we developed an on-line interactive flow mapping
service based on the spatiotemporal data cube model to effectively
represent the movement dynamics of groups of social media users (e.g., ILI
affected users) from a continent scale to a fine blocks level of scale.

The findings of this paper lay solid foundations for the future research in
spatiotemporal analysis of location-based social media data. On the
application side, the data cube model provides a novel structured
spatiotemporal data source that can be easily integrated with conventional
GIS and spatiotemporal analysis tools for mapping, modeling, and analyzing
large-scale complex spatiotemporal dynamics at different scales. In this
paper, we did not explicitly consider the structure of social network of
social media users, which conveys important interaction information between
social media users. People do things together with friends on a daily
basis, and interact with and get influenced by them. Valuable information
could be discovered for social media users by accounting for the activities
of their friends and their interaction with them. For example, social media
users identified as infected cases of ILI also put their frequently
interacted friends at high risks to get infected. Social medial data
provide access to the social interactions between friends, and thus make it
possible to investigate the spread of infection in a very fine individual
level. Another example is that we assumed the space-time trajectories are
step functions consisting of a sequence of moves between time-stamps and
locations. It is apparently not the case in reality. With the help of the
locations of close friends', we could estimate locations of social media
users at un-sampled time-stamps. The proposed data model is rather general
and flexible. In addition to the Twitter streams, data from other forms of
social media could also be incorporated. We are particularly interested in
the integration with the Foursquare data, which make senses of the
geographic coordinates with names and attributes of locations (e.g.,
restaurants, hospitals) and thus provide more detail information about
daily activities. How to deal with uncertainty of the location-based
social media data in the proposed data model is another topic that warrants
further investigation. In location-based social media data and the
data model, all of the spatiotemporal information, social media contents,
and the data mining results of contents are uncertain in reality.
Characterization of such uncertainties is desirable for effective use of
such data sources and the data model.

\section{Acknowledgments}
This material is based in part upon work supported by NSF under Grant
Numbers: \textit{0846655}, \textit{1047916}, and \textit{1354329}. This
work used the Extreme Science and Engineering Discovery Environment
(XSEDE), which is supported by National Science Foundation grant number
\textit{OCI-1053575}. Any opinions, findings, and conclusions or
recommendations expressed in this material are those of the authors and do
not necessarily reflect the views of the National Science Foundation. 

\bibliographystyle{model5-names}
\bibliography{main.bbl}

\begin{thebibliography}{46}
\expandafter\ifx\csname natexlab\endcsname\relax\def\natexlab#1{#1}\fi
\providecommand{\url}[1]{\texttt{#1}}
\providecommand{\href}[2]{#2}
\providecommand{\path}[1]{#1}
\providecommand{\DOIprefix}{doi:}
\providecommand{\ArXivprefix}{arXiv:}
\providecommand{\URLprefix}{URL: }
\providecommand{\Pubmedprefix}{pmid:}
\providecommand{\doi}[1]{\href{http://dx.doi.org/#1}{\path{#1}}}
\providecommand{\Pubmed}[1]{\href{pmid:#1}{\path{#1}}}
\providecommand{\bibinfo}[2]{#2}
\ifx\xfnm\relax \def\xfnm[#1]{\unskip,\space#1}\fi
\bibitem[{Andrienko et~al.(2012)Andrienko, Andrienko, Stange, Liebig \&
  Hecker}]{Andrienko2012}
\bibinfo{author}{Andrienko, N.}, \bibinfo{author}{Andrienko, G.},
  \bibinfo{author}{Stange, H.}, \bibinfo{author}{Liebig, T.}, \&
  \bibinfo{author}{Hecker, D.} (\bibinfo{year}{2012}).
\newblock \bibinfo{title}{Visual analytics for understanding spatial situations
  from episodic movement data}.
\newblock {\it \bibinfo{journal}{KI - Künstliche Intelligenz}\/},  {\it
  \bibinfo{volume}{26}\/}, \bibinfo{pages}{241--251}.
  \DOIprefix\doi{10.1007/s13218-012-0177-4}.
\bibitem[{Backstrom et~al.(2010)Backstrom, Sun \& Marlow}]{Backstrom2010}
\bibinfo{author}{Backstrom, L.}, \bibinfo{author}{Sun, E.}, \&
  \bibinfo{author}{Marlow, C.} (\bibinfo{year}{2010}).
\newblock \bibinfo{title}{Find me if you can: improving geographical prediction
  with social and spatial proximity}.
\newblock In {\it \bibinfo{booktitle}{Proceedings of the 19th international
  conference on World wide web}\/} (pp. \bibinfo{pages}{61--70}).
\newblock \bibinfo{organization}{ACM}.
\bibitem[{Bimonte et~al.(2010)Bimonte, Tchounikine, Miquel \&
  Pinet}]{Bimonte2010}
\bibinfo{author}{Bimonte, S.}, \bibinfo{author}{Tchounikine, A.},
  \bibinfo{author}{Miquel, M.}, \& \bibinfo{author}{Pinet, F.}
  (\bibinfo{year}{2010}).
\newblock \bibinfo{title}{When spatial analysis meets {OLAP}: Multidimensional
  model and operators}.
\newblock {\it \bibinfo{journal}{International Journal of Data Warehousing and
  Mining (IJDWM)}\/},  {\it \bibinfo{volume}{6}\/}, \bibinfo{pages}{33--60}.
\bibitem[{Burger et~al.(2011)Burger, Henderson, Kim \& Zarrella}]{Burger2011}
\bibinfo{author}{Burger, J.~D.}, \bibinfo{author}{Henderson, J.},
  \bibinfo{author}{Kim, G.}, \& \bibinfo{author}{Zarrella, G.}
  (\bibinfo{year}{2011}).
\newblock \bibinfo{title}{Discriminating gender on {T}witter}.
\newblock In {\it \bibinfo{booktitle}{Proceedings of the Conference on
  Empirical Methods in Natural Language Processing}\/} (pp.
  \bibinfo{pages}{1301--1309}).
\newblock \bibinfo{organization}{Association for Computational Linguistics}.
\bibitem[{Cranshaw et~al.(2012)Cranshaw, Schwartz, Hong \&
  Sadeh}]{Cranshaw2012}
\bibinfo{author}{Cranshaw, J.}, \bibinfo{author}{Schwartz, R.},
  \bibinfo{author}{Hong, J.}, \& \bibinfo{author}{Sadeh, N.}
  (\bibinfo{year}{2012}).
\newblock \bibinfo{title}{The livehoods project: Utilizing social media to
  understand the dynamics of a city}.
\newblock {\it \bibinfo{journal}{ICWSM'12}\/}, .
\bibitem[{Dobson et~al.(2000)Dobson, Coleman, Durfee \& Worley}]{Dobson2000}
\bibinfo{author}{Dobson, J.~E.}, \bibinfo{author}{Coleman, P.~R.},
  \bibinfo{author}{Durfee, R.~C.}, \& \bibinfo{author}{Worley, B.~A.}
  (\bibinfo{year}{2000}).
\newblock \bibinfo{title}{Landscan: a global population database for estimating
  populations at risk}.
\newblock {\it \bibinfo{journal}{Photogrammetric Engineering and Remote
  Sensing}\/},  {\it \bibinfo{volume}{66}\/}, \bibinfo{pages}{849--857}.
\bibitem[{Frank et~al.(2013)Frank, Mitchell, Dodds \& Danforth}]{Frank2013}
\bibinfo{author}{Frank, M.~R.}, \bibinfo{author}{Mitchell, L.},
  \bibinfo{author}{Dodds, P.~S.}, \& \bibinfo{author}{Danforth, C.~M.}
  (\bibinfo{year}{2013}).
\newblock \bibinfo{title}{Happiness and the patterns of life: A study of
  geolocated tweets}.
\newblock {\it \bibinfo{journal}{Scientific reports}\/},  {\it
  \bibinfo{volume}{3}\/}.
\bibitem[{Gao \& Liu(2013)}]{Gao2013}
\bibinfo{author}{Gao, H.}, \& \bibinfo{author}{Liu, H.} (\bibinfo{year}{2013}).
\newblock \bibinfo{title}{Data analysis on location-based social networks}.
\newblock In \bibinfo{editor}{A.~Chin}, \& \bibinfo{editor}{D.~Zhang} (Eds.),
  {\it \bibinfo{booktitle}{Mobile Social Networking: An Innovative Approach}\/}
  (pp. \bibinfo{pages}{165--194}).
\bibitem[{Golfarelli et~al.(1998)Golfarelli, Maio \& Rizzi}]{Golfarelli1998}
\bibinfo{author}{Golfarelli, M.}, \bibinfo{author}{Maio, D.}, \&
  \bibinfo{author}{Rizzi, S.} (\bibinfo{year}{1998}).
\newblock \bibinfo{title}{The dimensional fact model: A conceptual model for
  data warehouses}.
\newblock {\it \bibinfo{journal}{International Journal of Cooperative
  Information Systems}\/},  {\it \bibinfo{volume}{7}\/},
  \bibinfo{pages}{215--247}.
\bibitem[{Gonzalez et~al.(2008)Gonzalez, Hidalgo \& Barabasi}]{Gonzalez2008a}
\bibinfo{author}{Gonzalez, M.~C.}, \bibinfo{author}{Hidalgo, C.~A.}, \&
  \bibinfo{author}{Barabasi, A.-L.} (\bibinfo{year}{2008}).
\newblock \bibinfo{title}{Understanding individual human mobility patterns}.
\newblock {\it \bibinfo{journal}{Nature}\/},  {\it \bibinfo{volume}{453}\/},
  \bibinfo{pages}{779--782}.
\bibitem[{Goodchild(2004)}]{Goodchild2004d}
\bibinfo{author}{Goodchild, M.} (\bibinfo{year}{2004}).
\newblock \bibinfo{title}{{GIScience, geography, form, and process}}.
\newblock {\it \bibinfo{journal}{Annals of the Association of American}\/},
  {\it \bibinfo{volume}{94}\/}, \bibinfo{pages}{709--714}.
\bibitem[{Gray(2009)}]{Gray2009}
\bibinfo{author}{Gray, J.} (\bibinfo{year}{2009}).
\newblock \bibinfo{title}{{J}im {G}ray on e{S}cience: a transformed scientific
  method}.
\newblock {\it \bibinfo{journal}{The fourth paradigm: Data-intensive scientific
  discovery}\/}, .
\bibitem[{Gray et~al.(1997)Gray, Chaudhuri, Bosworth, Layman, Reichart,
  Venkatrao, Pellow \& Pirahesh}]{Gray1997}
\bibinfo{author}{Gray, J.}, \bibinfo{author}{Chaudhuri, S.},
  \bibinfo{author}{Bosworth, A.}, \bibinfo{author}{Layman, A.},
  \bibinfo{author}{Reichart, D.}, \bibinfo{author}{Venkatrao, M.},
  \bibinfo{author}{Pellow, F.}, \& \bibinfo{author}{Pirahesh, H.}
  (\bibinfo{year}{1997}).
\newblock \bibinfo{title}{Data cube: A relational aggregation operator
  generalizing group-by, cross-tab, and sub-totals}.
\newblock {\it \bibinfo{journal}{Data Mining and Knowledge Discovery}\/},  {\it
  \bibinfo{volume}{1}\/}, \bibinfo{pages}{29--53}.
\bibitem[{Guo et~al.(2006)Guo, Chen, MacEachren \& Liao}]{Guo2006}
\bibinfo{author}{Guo, D.}, \bibinfo{author}{Chen, J.},
  \bibinfo{author}{MacEachren, A.~M.}, \& \bibinfo{author}{Liao, K.}
  (\bibinfo{year}{2006}).
\newblock \bibinfo{title}{A visualization system for space-time and
  multivariate patterns (vis-stamp)}.
\newblock {\it \bibinfo{journal}{Visualization and Computer Graphics, IEEE
  Transactions on}\/},  {\it \bibinfo{volume}{12}\/},
  \bibinfo{pages}{1461--1474}.
\bibitem[{G{\"u}ting \& Schneider(2005)}]{Guting2005}
\bibinfo{author}{G{\"u}ting, R.~H.}, \& \bibinfo{author}{Schneider, M.}
  (\bibinfo{year}{2005}).
\newblock {\it \bibinfo{title}{Moving Objects Databases}\/}.
\newblock \bibinfo{publisher}{Morgan Kaufmann}.
\bibitem[{H{\"a}gerstraand(1970)}]{Hagerstrand1970}
\bibinfo{author}{H{\"a}gerstraand, T.} (\bibinfo{year}{1970}).
\newblock \bibinfo{title}{What about people in regional science?}
\newblock {\it \bibinfo{journal}{Papers in regional science}\/},  {\it
  \bibinfo{volume}{24}\/}, \bibinfo{pages}{7--24}.
\bibitem[{Han et~al.(1998)Han, Stefanovic \& Koperski}]{HSK98}
\bibinfo{author}{Han, J.}, \bibinfo{author}{Stefanovic, N.}, \&
  \bibinfo{author}{Koperski, K.} (\bibinfo{year}{1998}).
\newblock \bibinfo{title}{{Selective Materialization: An Efficient Method for
  Spatial Data Cube Construction}}.
\newblock In {\it \bibinfo{booktitle}{Proc. 1998 Pacific-Asia Conf. Knowledge
  Discovery and Data Mining (PAKDD'98)}\/}.
\newblock \bibinfo{address}{Melbourne, Australia}.
\bibitem[{Holten \& Van~Wijk(2009)}]{Holten2009}
\bibinfo{author}{Holten, D.}, \& \bibinfo{author}{Van~Wijk, J.~J.}
  (\bibinfo{year}{2009}).
\newblock \bibinfo{title}{Force-directed edge bundling for graph
  visualization}.
\newblock In {\it \bibinfo{booktitle}{Computer Graphics Forum}\/} (pp.
  \bibinfo{pages}{983--990}).
\newblock \bibinfo{organization}{Wiley Online Library}
  volume~\bibinfo{volume}{28}.
\bibitem[{Inmon(2005)}]{Inmon2005}
\bibinfo{author}{Inmon, W.~H.} (\bibinfo{year}{2005}).
\newblock {\it \bibinfo{title}{Building the Data Warehouse}\/}.
\newblock \bibinfo{publisher}{John Wiley \& Sons}.
\bibitem[{Center~for International Earth Science Information
  Network~(CIESIN)(2004)}]{CIESIN2004}
\bibinfo{author}{Center~for International Earth Science Information
  Network~(CIESIN), .} (\bibinfo{year}{2004}).
\newblock {\it \bibinfo{title}{Gridded Population of the World (GPW),
  ver.3}\/}.
\bibitem[{Kaplan \& Haenlein(2010)}]{kaplan2010}
\bibinfo{author}{Kaplan, A.}, \& \bibinfo{author}{Haenlein, M.}
  (\bibinfo{year}{2010}).
\newblock \bibinfo{title}{{Users of the world, unite! The challenges and
  opportunities of Social Media}}.
\newblock {\it \bibinfo{journal}{Business horizons}\/},  {\it
  \bibinfo{volume}{53}\/}, \bibinfo{pages}{59--68}.
\bibitem[{Leonardi et~al.(2014)Leonardi, Orlando, Raffaet{\`a}, Roncato,
  Silvestri, Andrienko \& Andrienko}]{Leonardi2014}
\bibinfo{author}{Leonardi, L.}, \bibinfo{author}{Orlando, S.},
  \bibinfo{author}{Raffaet{\`a}, A.}, \bibinfo{author}{Roncato, A.},
  \bibinfo{author}{Silvestri, C.}, \bibinfo{author}{Andrienko, G.}, \&
  \bibinfo{author}{Andrienko, N.} (\bibinfo{year}{2014}).
\newblock \bibinfo{title}{A general framework for trajectory data warehousing
  and visual olap}.
\newblock {\it \bibinfo{journal}{GeoInformatica}\/},  {\it
  \bibinfo{volume}{18}\/}, \bibinfo{pages}{273--312}.
\bibitem[{Li \& Goodchild(2012)}]{Li2012b}
\bibinfo{author}{Li, L.}, \& \bibinfo{author}{Goodchild, M.~F.}
  (\bibinfo{year}{2012}).
\newblock \bibinfo{title}{Constructing places from spatial footprints}.
\newblock In {\it \bibinfo{booktitle}{Proceedings of the 1st ACM SIGSPATIAL
  International Workshop on Crowdsourced and Volunteered Geographic
  Information}\/} (pp. \bibinfo{pages}{15--21}).
\newblock \bibinfo{organization}{ACM}.
\bibitem[{Lins et~al.(2013)Lins, Klosowski \& Scheidegger}]{Lins2013}
\bibinfo{author}{Lins, L.}, \bibinfo{author}{Klosowski, J.~T.}, \&
  \bibinfo{author}{Scheidegger, C.} (\bibinfo{year}{2013}).
\newblock \bibinfo{title}{Nanocubes for real-time exploration of spatiotemporal
  datasets}.
\newblock {\it \bibinfo{journal}{Visualization and Computer Graphics, IEEE
  Transactions on}\/},  {\it \bibinfo{volume}{19}\/},
  \bibinfo{pages}{2456--2465}.
\bibitem[{Manyika et~al.(2011)Manyika, Chui, Brown, Bughin, Dobbs, Roxburgh \&
  Byers}]{Manyika2011}
\bibinfo{author}{Manyika, J.}, \bibinfo{author}{Chui, M.},
  \bibinfo{author}{Brown, B.}, \bibinfo{author}{Bughin, J.},
  \bibinfo{author}{Dobbs, R.}, \bibinfo{author}{Roxburgh, C.}, \&
  \bibinfo{author}{Byers, A.~H.} (\bibinfo{year}{2011}).
\newblock \bibinfo{title}{Big data: The next frontier for innovation,
  competition, and productivity}.
\newblock {\it \bibinfo{journal}{McKinsey Global Institute}\/},  (pp.
  \bibinfo{pages}{1--137}).
\bibitem[{Morstatter et~al.(2013)Morstatter, Pfeffer, Liu \&
  Carley}]{Morstatter2013}
\bibinfo{author}{Morstatter, F.}, \bibinfo{author}{Pfeffer, J.},
  \bibinfo{author}{Liu, H.}, \& \bibinfo{author}{Carley, K.~M.}
  (\bibinfo{year}{2013}).
\newblock \bibinfo{title}{Is the sample good enough? {C}omparing data from
  {T}witter’s streaming {API} with {T}witter’s {F}irehose}.
\newblock {\it \bibinfo{journal}{Proceedings of ICWSM}\/}, .
\bibitem[{Nagel et~al.(2013)Nagel, Tsou, Spitzberg, An, Gawron, Gupta, Yang,
  Han, Peddecord, Lindsay et~al.}]{Nagel2013}
\bibinfo{author}{Nagel, A.~C.}, \bibinfo{author}{Tsou, M.-H.},
  \bibinfo{author}{Spitzberg, B.~H.}, \bibinfo{author}{An, L.},
  \bibinfo{author}{Gawron, J.~M.}, \bibinfo{author}{Gupta, D.~K.},
  \bibinfo{author}{Yang, J.-A.}, \bibinfo{author}{Han, S.},
  \bibinfo{author}{Peddecord, K.~M.}, \bibinfo{author}{Lindsay, S.} et~al.
  (\bibinfo{year}{2013}).
\newblock \bibinfo{title}{The complex relationship of realspace events and
  messages in cyberspace: Case study of influenza and pertussis using tweets}.
\newblock {\it \bibinfo{journal}{Journal of medical Internet research}\/},
  {\it \bibinfo{volume}{15}\/}.
\bibitem[{O'Connor \& Balasubramanyan(2010)}]{OConnor2010}
\bibinfo{author}{O'Connor, B.}, \& \bibinfo{author}{Balasubramanyan, R.}
  (\bibinfo{year}{2010}).
\newblock \bibinfo{title}{{From {T}weets to {P}olls: Linking text sentiment to
  public opinion time series}}.
\newblock In {\it \bibinfo{booktitle}{Proceedings of the Fourth International
  AAAI Conference on Weblogs and Social Media}\/}.
\bibitem[{Openshaw(1983)}]{OpenShaw1983}
\bibinfo{author}{Openshaw, S.} (\bibinfo{year}{1983}).
\newblock {\it \bibinfo{title}{The Modifiable Areal Unit Problem}\/}
  volume~\bibinfo{volume}{38}.
\newblock \bibinfo{publisher}{Geo books Norwich}.
\bibitem[{Orlando et~al.(2007)Orlando, Orsini, Raffaet{\`a}, Roncato \&
  Silvestri}]{Orlando2007}
\bibinfo{author}{Orlando, S.}, \bibinfo{author}{Orsini, R.},
  \bibinfo{author}{Raffaet{\`a}, A.}, \bibinfo{author}{Roncato, A.}, \&
  \bibinfo{author}{Silvestri, C.} (\bibinfo{year}{2007}).
\newblock \bibinfo{title}{Trajectory data warehouses: Design and implementation
  issues.}, .
\bibitem[{Papadias et~al.(2002)Papadias, Tao, Kanis \& Zhang}]{Papadias2002}
\bibinfo{author}{Papadias, D.}, \bibinfo{author}{Tao, Y.},
  \bibinfo{author}{Kanis, P.}, \& \bibinfo{author}{Zhang, J.}
  (\bibinfo{year}{2002}).
\newblock \bibinfo{title}{Indexing spatio-temporal data warehouses}.
\newblock In {\it \bibinfo{booktitle}{Data Engineering, 2002. Proceedings. 18th
  International Conference on}\/} (pp. \bibinfo{pages}{166--175}).
\newblock \bibinfo{organization}{IEEE}.
\bibitem[{Park et~al.(2013)Park, Yu, Park \& Kim}]{Park2013}
\bibinfo{author}{Park, D.}, \bibinfo{author}{Yu, J.}, \bibinfo{author}{Park,
  J.-S.}, \& \bibinfo{author}{Kim, M.-S.} (\bibinfo{year}{2013}).
\newblock \bibinfo{title}{Netcube: a comprehensive network traffic analysis
  model based on multidimensional olap data cube}.
\newblock {\it \bibinfo{journal}{International Journal of Network
  Management}\/},  {\it \bibinfo{volume}{23}\/}, \bibinfo{pages}{101--118}.
\bibitem[{Rao et~al.(2010)Rao, Yarowsky, Shreevats \& Gupta}]{Rao2010}
\bibinfo{author}{Rao, D.}, \bibinfo{author}{Yarowsky, D.},
  \bibinfo{author}{Shreevats, A.}, \& \bibinfo{author}{Gupta, M.}
  (\bibinfo{year}{2010}).
\newblock \bibinfo{title}{Classifying latent user attributes in {T}witter}.
\newblock In {\it \bibinfo{booktitle}{Proceedings of the 2nd international
  workshop on Search and mining user-generated contents}\/} (pp.
  \bibinfo{pages}{37--44}).
\newblock \bibinfo{organization}{ACM}.
\bibitem[{Sadilek \& Krumm(2012)}]{Sadilek2012a}
\bibinfo{author}{Sadilek, A.}, \& \bibinfo{author}{Krumm, J.}
  (\bibinfo{year}{2012}).
\newblock \bibinfo{title}{Far out: {P}redicting long-term human mobility}.
\newblock In {\it \bibinfo{booktitle}{Proceedings of the Twenty-Sixth AAAI
  Conference on Artificial Intelligence}\/} (pp. \bibinfo{pages}{814--820}).
\bibitem[{Shekhar et~al.(2001)Shekhar, Lu, Tan, Chawla \&
  Vatsavai}]{Shekhar2001}
\bibinfo{author}{Shekhar, S.}, \bibinfo{author}{Lu, C.-T.},
  \bibinfo{author}{Tan, X.}, \bibinfo{author}{Chawla, S.}, \&
  \bibinfo{author}{Vatsavai, R.~R.} (\bibinfo{year}{2001}).
\newblock \bibinfo{title}{{Map Cube: A Visualization Tool for Spatial Data
  Warehouses}}.
\newblock In \bibinfo{editor}{H.~J. Miller}, \& \bibinfo{editor}{J.~Han}
  (Eds.), {\it \bibinfo{booktitle}{Geographic Data Mining and Knowledge
  Discovery}\/} (pp. \bibinfo{pages}{73--108}).
\newblock \bibinfo{publisher}{Taylor and Francis}.
\bibitem[{Signorini et~al.(2011)Signorini, Segre \& Polgreen}]{Signorini2011}
\bibinfo{author}{Signorini, A.}, \bibinfo{author}{Segre, A.~M.}, \&
  \bibinfo{author}{Polgreen, P.~M.} (\bibinfo{year}{2011}).
\newblock \bibinfo{title}{The use of {T}witter to track levels of disease
  activity and public concern in the us during the influenza a h1n1 pandemic}.
\newblock {\it \bibinfo{journal}{PloS one}\/},  {\it \bibinfo{volume}{6}\/},
  \bibinfo{pages}{e19467}.
\bibitem[{Tang et~al.(2012)Tang, Yu, Kim, Han, Peng, Sun, Leung \&
  La~Porta}]{Tang2012}
\bibinfo{author}{Tang, L.-A.}, \bibinfo{author}{Yu, X.}, \bibinfo{author}{Kim,
  S.}, \bibinfo{author}{Han, J.}, \bibinfo{author}{Peng, W.-C.},
  \bibinfo{author}{Sun, Y.}, \bibinfo{author}{Leung, A.}, \&
  \bibinfo{author}{La~Porta, T.} (\bibinfo{year}{2012}).
\newblock \bibinfo{title}{Multidimensional sensor data analysis in
  cyber-physical system: an atypical cube approach}.
\newblock {\it \bibinfo{journal}{International Journal of Distributed Sensor
  Networks}\/},  {\it \bibinfo{volume}{2012}\/}.
\bibitem[{Tao et~al.(2004)Tao, Kollios, Considine, Li \& Papadias}]{Tao2004}
\bibinfo{author}{Tao, Y.}, \bibinfo{author}{Kollios, G.},
  \bibinfo{author}{Considine, J.}, \bibinfo{author}{Li, F.}, \&
  \bibinfo{author}{Papadias, D.} (\bibinfo{year}{2004}).
\newblock \bibinfo{title}{Spatio-temporal aggregation using sketches}.
\newblock In {\it \bibinfo{booktitle}{Data Engineering, 2004. Proceedings. 20th
  International Conference on}\/} (pp. \bibinfo{pages}{214--225}).
\newblock \bibinfo{organization}{IEEE}.
\bibitem[{Tobler(1987)}]{Tobler1987}
\bibinfo{author}{Tobler, W.~R.} (\bibinfo{year}{1987}).
\newblock \bibinfo{title}{Experiments in migration mapping by computer}.
\newblock {\it \bibinfo{journal}{The American Cartographer}\/},  {\it
  \bibinfo{volume}{14}\/}.
\bibitem[{Tsou \& Leitner(2013)}]{Tsou2013}
\bibinfo{author}{Tsou, M.-H.}, \& \bibinfo{author}{Leitner, M.}
  (\bibinfo{year}{2013}).
\newblock \bibinfo{title}{Visualization of social media: seeing a mirage or a
  message?}
\newblock {\it \bibinfo{journal}{Cartography and Geographic Information
  Science}\/},  {\it \bibinfo{volume}{40}\/}, \bibinfo{pages}{55--60}.
\bibitem[{Verbeek et~al.(2011)Verbeek, Buchin \& Speckmann}]{Verbeek2011}
\bibinfo{author}{Verbeek, K.}, \bibinfo{author}{Buchin, K.}, \&
  \bibinfo{author}{Speckmann, B.} (\bibinfo{year}{2011}).
\newblock \bibinfo{title}{Flow map layout via spiral trees}.
\newblock {\it \bibinfo{journal}{IEEE Transactions on Visualization and
  Computer Graphics}\/},  {\it \bibinfo{volume}{17}\/},
  \bibinfo{pages}{2536--2544}.
\bibitem[{Wang(2010)}]{Wang2010}
\bibinfo{author}{Wang, S.} (\bibinfo{year}{2010}).
\newblock \bibinfo{title}{{A {CyberGIS} framework for the synthesis of
  cyberinfrastructure, GIS, and spatial analysis}}.
\newblock {\it \bibinfo{journal}{Annals of the Association of American
  Geographers}\/},  {\it \bibinfo{volume}{100}\/}, \bibinfo{pages}{535--557}.
\bibitem[{Wang et~al.(2013)Wang, Cao, Zhang \& Zhao}]{Wang2013}
\bibinfo{author}{Wang, S.}, \bibinfo{author}{Cao, G.}, \bibinfo{author}{Zhang,
  Z.}, \& \bibinfo{author}{Zhao, Y.} (\bibinfo{year}{2013}).
\newblock \bibinfo{title}{A {CyberGIS} environment for analysis of
  location-based social media data}.
\newblock {\it \bibinfo{journal}{Advanced Location-based Technologies and
  Services}\/},  (p. \bibinfo{pages}{187}).
\bibitem[{Wright \& Wang(2011)}]{Wright2011}
\bibinfo{author}{Wright, D.}, \& \bibinfo{author}{Wang, S.}
  (\bibinfo{year}{2011}).
\newblock \bibinfo{title}{{The emergence of spatial cyberinfrastructure}}.
\newblock {\it \bibinfo{journal}{Proceedings of the National Academy of
  Sciences}\/},  {\it \bibinfo{volume}{108}\/}, \bibinfo{pages}{5488--5491}.
\bibitem[{Wu et~al.(2014)Wu, Zhi, Sui \& Liu}]{Wu2014}
\bibinfo{author}{Wu, L.}, \bibinfo{author}{Zhi, Y.}, \bibinfo{author}{Sui, Z.},
  \& \bibinfo{author}{Liu, Y.} (\bibinfo{year}{2014}).
\newblock \bibinfo{title}{Intra-urban human mobility and activity transition:
  Evidence from social media check-in data}.
\newblock {\it \bibinfo{journal}{PloS ONE}\/},  {\it \bibinfo{volume}{9}\/},
  \bibinfo{pages}{e97010}.
\bibitem[{Zheng \& Zhou(2011)}]{Zheng2011}
\bibinfo{author}{Zheng, Y.}, \& \bibinfo{author}{Zhou, X.}
  (\bibinfo{year}{2011}).
\newblock {\it \bibinfo{title}{{Computing with Spatial Trajectories}}\/}.
\newblock \bibinfo{publisher}{Springer}.

\end{thebibliography}
\end{document}